\documentstyle[osa,graphicx]{revtex}

\begin{document}

\title{All Optical Quantum Teleportation}
\author{T.C.Ralph}
\address{Department of Physics, Faculty of
Science, \\ The Australian National University, \\ ACT 0200 Australia \\ 
Fax: +61 6 249 0741  Telephone: +61 6 249 4105 \\ E-mail: 
Timothy.Ralph@anu.edu.au}
\maketitle

\begin{center}
\scriptsize (12th November 1998)
\end{center}

\begin{abstract}

We propose an all optical, continuous variable, quantum teleportation 
scheme based on optical parametric amplifiers. 

\end{abstract}

\vspace{10 mm}

Quantum teleportation \cite{ben93} is a technique via which the 
quantum state of a system can be transmitted to a distant station 
through the direct transmission of only classical information. This 
remarkable effect is made possible by the sharing of a non-locally 
entangled state by the sender and receiver, constituting an indirect 
quantum channel. The original proposals were for the teleportation of 
single particles and experimental demonstrations with photons
have been made, albeit 
with very low efficiencies \cite{bou97,bos98}. More recently 
continuous variable schemes have been proposed 
\cite{vai94,bra981,ral981} which can teleport continuous multi-particle 
states. A high efficiency, experimental demonstration of the 
teleportation of a single mode, coherent optical field has been made 
\cite{fur98}.

Teleportation has applications in quantum computing \cite{bras98} and 
general quantum information manipulation \cite{ben95}. Continuous 
variable teleportation can be applied both through continuous variable 
generalizations of discrete manipulations \cite{bra982} and 
continuous variable manipulations of discrete properties 
\cite{pol98}. Teleportation of optical fields holds great promise due 
to the power of the required optical tools and the maturity of 
relevant optical communications technology. 

One drawback of the current schemes is the electro-optic 
manipulation of the classical information \cite{fur98,ral981}. 
The sender photo-detects 
the incoming field, sends the information electronically and the 
receiver reconstructs the state through electro-optic modulation. This 
imposes severe limitations on the bandwidth of the information that 
can be successfully teleported. All optical manipulation of the 
classical information would clearly broaden the potential 
applications. In this letter we propose such an all optical 
teleportation system which employs currently available technology.

Consider the ``classical teleportation'' schemes shown in Fig.1. 
In Fig.1(a) an idealized electro-optic scheme is depicted. The 
input light mode, with annihilation operator $a_{in}$, 
is split at a 50:50 beamsplitter. The 
in-phase quadrature ($X_{j}^{+}=a_{j}+a_{j}^{\dagger}$) of one beam 
and the out-of-phase quadrature ($X_{j}^{-}=i(a_{j}-a_{j}^{\dagger})$) 
of the other beam, are measured using ideal homodyne detectors and 
local oscillators (LO). The 
photocurrents of the in-phase and out-of-phase detectors 
are given by $A_{c}^{+}=K/2(a_{in}+a_{in}^{\dagger}+
v_{1}+v_{1}^{\dagger})$ and $A_{c}^{-}=iK/2(a_{in}-a_{in}^{\dagger}-
v_{1}+v_{1}^{\dagger})$ respectively. Here $K$ is a constant of proportionality 
and $v_{1}$ is the vacuum mode entering at the unused port of the 
beamsplitter. The photo-currents can be combined to give the 
classical channel 
\begin{eqnarray}
A_{c} & = & A_{c}^{+}-iA_{c}^{-}\nonumber\\
 & = & K(a_{in}+v_{1}^{\dagger})
\label{Ac}
\end{eqnarray}
This photo-current is sent to the receiver 
who uses it to ``displace'' the vacuum input $v_{2}$ via a standard 
Mach-Zender arrangement (note that a simpler direct modulation 
technique may be used for bright beams \cite{ral981}). 
The output field is $a_{out}=\lambda 
A_{c}-v_{2}$ where $\lambda$ is an adjustable gain proportional to the 
local oscillator intensity. If we set 
$\lambda \times K=1$ we obtain the output field
\begin{equation}
a_{out}=a_{in}+v_{1}^{\dagger}-v_{2}
\label{out}
\end{equation}
The input is retrieved but is polluted by two independent vacuum modes. The 
presence of these vacuum modes is unavoidable and reduces both the information 
transferred from the input to the output beam 
and the correlation between the input 
and output beams \cite{ral981}. The result given by Eq.~\ref{out} sets 
an ideal limit to the similarity of the input and output modes for 
classical teleportation. However in practice, due to the limitations 
of detectors, electronic amplifiers and modulators, Eq.~\ref{out} will 
only be approximately satisfied for a narrow range of RF frequencies.

An all-optical analogue of the preceding scheme is shown in 
Fig.1(b). Now the input light mode, $a_{in}$, is sent through a linear optical 
amplifier by the sender. For an ideal linear amplifier the output will be 
given by \cite{cav81} 
\begin{equation}
a_{c}=\sqrt{G}a_{in}+\sqrt{G-1}v_{1}^{\dagger}
\label{ac}
\end{equation}
where $G$ is the amplifier gain and $v_{1}$ is a vacuum noise input. 
If the gain is sufficiently large ($G>>1$) then $a_{c}$ can be 
regarded as a classical field. This is because the conjugate quadrature variables 
$X_{c}^{+}=a_{c}+a_{c}^{\dagger}$ and 
$X_{c}^{-}=i(a_{c}-a_{c}^{\dagger}$) both have uncertainties much 
greater than the quantum limit, i.e. $\Delta (X_{c}^{\pm})^{2}>>1$.  
This means that simultaneous measurements of the 
conjugate quadratures can extract 
all the information carried by $a_{c}$ with negligible penalty. The quantum 
noise added due to the simultaneous measurements will be negligible 
compared to the amplified quadrature uncertainties. 
Similarly, $a_{c}$ can suffer propagation loss and compensating 
linear amplification without degradation. 
Hence, effectively, this is a classical channel. Notice that in the 
limit $G>>1$ Eq.~\ref{Ac} and Eq.~\ref{ac} are identical under the 
substitution $K=\sqrt{G}$.

The receiver attempts 
to retrieve the quantum state of the input by simply attenuating the 
the beam with a beamsplitter of 
transmission $\varepsilon$. The output field is 
$a_{out}=\sqrt{\varepsilon}a_{c}-\sqrt{1-\varepsilon}v_{2}$ where $v_{2}$ is the 
vacuum mode incident on the unused port of the beamsplitter. If we 
set $\varepsilon=1/G$ we obtain
\begin{eqnarray}
a_{out} & = & a_{in}+\sqrt{{G-1}\over{G}}(v_{1}^{\dagger}-v_{2})\nonumber\\
 & \approx & a_{in}+v_{1}^{\dagger}-v_{2}
\label{sn}
\end{eqnarray}
where the approximate equality is for $G>>1$. This is identical to 
the result of Eq.~\ref{out} but here it has been achieved using only 
beamsplitters and a linear amplifer, both of which can have a flat response 
over a large range of frequencies. 

The current proposals and demonstrations of quantum teleportation use 
the basic set-up of Fig.1(a) but replace the independent vacuums, 
$v_{1}$ and $v_{2}$ with non-locally entangled inputs. From the 
formal equivalence of Figs.1(a) and 1(b) it is 
clear that a similar trick should be possible with the all-optical scheme.
Hence consider the all optical ``quantum teleportation'' scheme shown in 
Fig.2. The arrangement is the same as for the all-optical classical scheme 
except that now the two vacuum inputs have been replaced by Einstein, 
Podolsky, Rosen (EPR) entangled beams \cite{ein35}, 
$b_{1}$ and $b_{2}$. Such beams 
have the very strong correlation property that both their difference 
amplitude quadrature variance,
$\Delta(X_{b1}^{+}-X_{b2}^{+})^{2}$, and their sum phase quadrature 
variance, $\Delta(X_{b1}^{-}+X_{b2}^{-})^{2}$, are less than the 
quantum limit (=1). Such beams can be generated by non-degenerate 
parametric amplification \cite{ou92} or by the mixing of independent squeezed 
sources \cite{ral981}. For non-degenerate parametric amplification 
these beams can be represented by
\begin{eqnarray}
b_{1} & = & \sqrt{H}v_{1}+\sqrt{H-1}v_{2}^{\dagger}\nonumber\\
b_{2} & = & \sqrt{H}v_{2}+\sqrt{H-1}v_{1}^{\dagger}
\end{eqnarray}
where $H$ is the parametric gain and $v_{1}$ and $v_{2}$ are vacuum 
inputs. The calculation proceeds formally as before, with the 
amplified field still clearly a classical field, and the output of 
the beamsplitter (with $\varepsilon=1/G$) given by
\begin{equation}
a_{out} \approx a_{in}+b_{1}^{\dagger}-b_{2}
\end{equation}
However now, because of the EPR correlation between $b_{1}$ and $b_{2}$, 
this reduces to
\begin{equation}
a_{out} \approx a_{in}+(\sqrt{H}-\sqrt{H-1})(v_{1}^{\dagger}-v_{2})
\end{equation}
and in the limit of very high parametric gain ($H \to \infty$) the 
output becomes identical to the input ($a_{out} \to a_{in}$). This is 
quantum teleportation as the only direct link between the input and 
output is the classical field $a_{c}$, yet perfect reconstruction of 
the input state is, in principle, possible with a sufficiently strong EPR 
correlation. The uncertainty principle is not compromised because the 
variances of each of the quadratures of $b_{1}$ by themselves are very noisy. 
Thus the information about $a_{in}$ carried on the classical field is 
buried in this noise and cannot be extracted by looking at the 
classical field alone.

The question remains as to how the linear amplifier in Fig.2 could 
be constructed. This is not trivial as in standard optical amplifiers 
the source of the vacuum mode is not available for modification. 
For example, in a laser amplifier the physical origin 
of the vacuum input is collisionally or phonon induced dipole 
fluctuations of the gain medium \cite{yam86}. One solution is shown 
schematically in Fig.3. The input beam is mixed with the EPR beam, 
$b_{1}$, at a 50:50 beamsplitter. The output beams are
\begin{eqnarray}
c & = & {{1}\over{\sqrt{2}}}(a_{in}+b_{1})\nonumber\\
d & = & {{1}\over{\sqrt{2}}}(a_{in}-b_{1})
\end{eqnarray}
The beams are amplified by degenerate parametric 
amplifiers of equal gains but with a $\pi$ phase shift between there 
pump ($E$) phases. This results in the outputs
\begin{eqnarray}
c' & = & \sqrt{G}c+\sqrt{G-1}c^{\dagger}\nonumber\\
d' & = & \sqrt{G}d-\sqrt{G-1}d^{\dagger}
\end{eqnarray}
Recombining these beams on a beamsplitter then produces the desired 
output: $a_{c}=\sqrt{G}a_{in}+\sqrt{G-1}b_{1}^{\dagger}$.

The importance of these results is both practical and fundamental. As 
we noted in the introduction, all-optical transitions relax the 
bandwidth limitations inherent in opto-electronic manipulations. In 
fact parametric amplification has such a broad gain profile 
\cite{tsu98} that 
one could envisage teleporting multiple-temporal-mode fields such as 
optical pulses with this set-up. This would be completely impractical 
using electro-optic means. The ability to perform optical 
manipulations on the classical channel may be important for quantum 
information applications. The experimental set-up depicted in Figs.2 
and 3, though not trivial, is clearly within the reach of present 
technology. Its greater complexity over an electro-optic set-up 
may be warranted by its improved versatility.

From a more fundamental point of view, this system illustrates 
explicitly that teleportation does not involve intrinsically irreversible 
processes \cite{bra96}. Beam-splitters and parametric amplifiers are 
time symmetric devices unlike the detectors and modulators in the 
standard schemes. The irreversibility of the entire process comes from 
the wasting of information, firstly in the amplification process 
when beam $e$ is discarded (see Fig.3) and secondly in the 
reconstruction process when beam $f$ is discarded (see Fig.2). Further 
study of all-optical systems like this one may bring new insights into the 
fundamental processes at work.

We thank C.M.Savage for useful discussions.
T.C.Ralph is an ARC Postdoctoral Fellow.

\begin{figure}
\includegraphics[width=10cm]{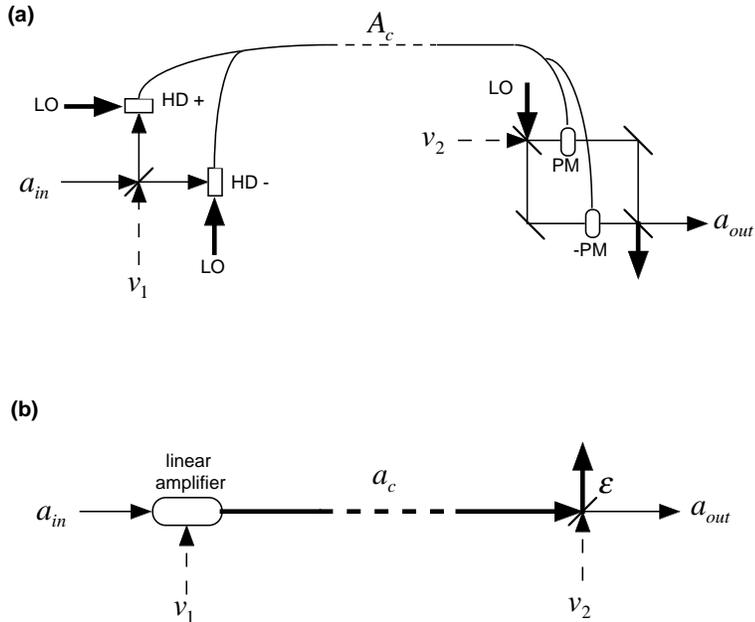}
\label{fig1}
 \caption{Schematics of classical teleportation; (a) 
 electro-optic and (b) all-optical. The following symbols are 
 used:LO $\equiv$ Local oscillator; HD+ $\equiv$ homodyne detection 
 of the amplitude quadrature; HD- $\equiv$ homodyne detection 
 of the phase quadrature; PM $\equiv$ phase modulation; -PM $\equiv$ 
 phase modulation with a $\pi$ phase shift.}
\end{figure}

\begin{figure}
\includegraphics[width=10cm]{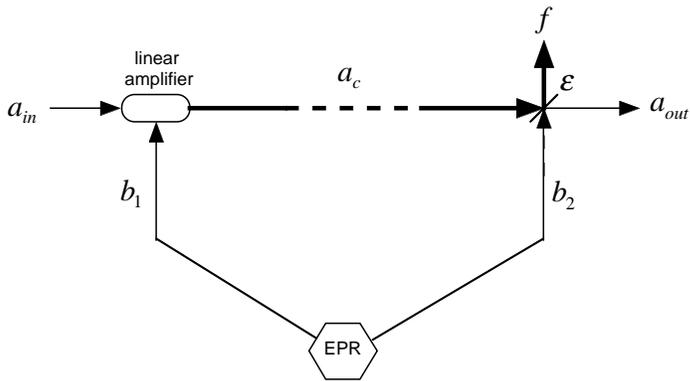}
\label{fig2}
 \caption{Schematic of all-optical quantum teleportation.}
\end{figure}

\begin{figure}
\includegraphics[width=10cm]{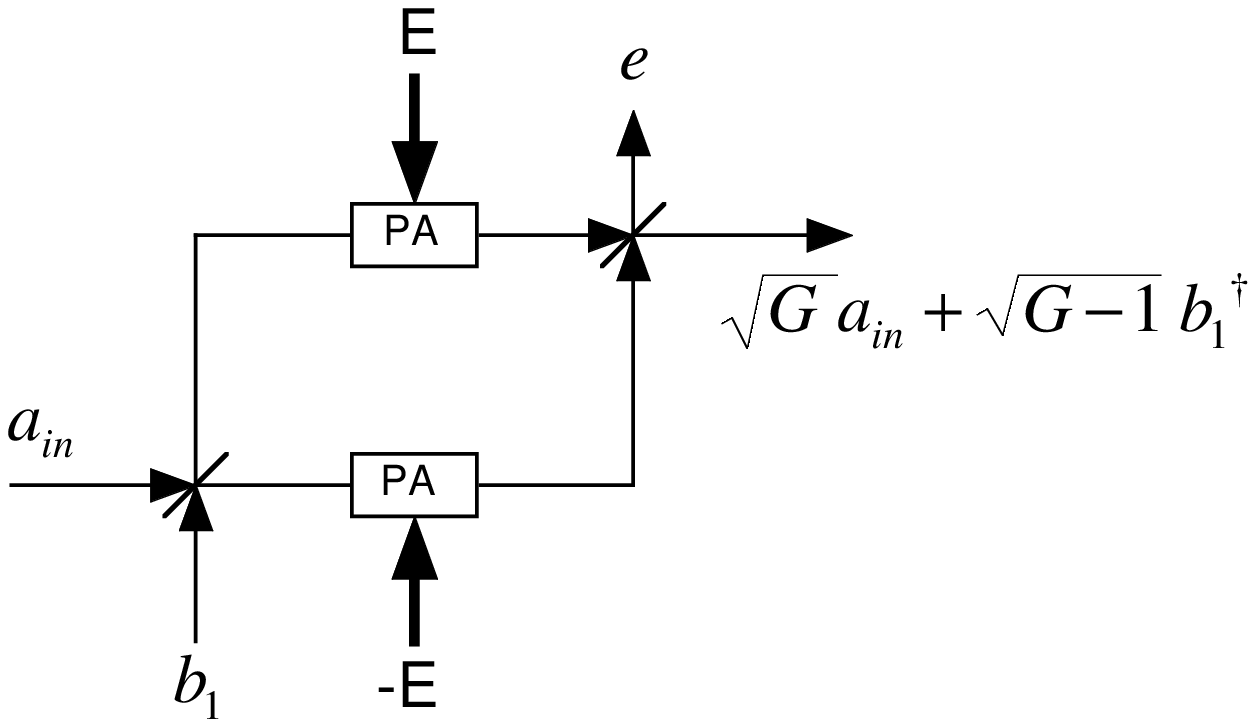}
\label{fig3}
 \caption{Schematic of linear amplifier with controllable noise penalty. 
 The following symbols are used: PA $\equiv$ parametric amplifier; E 
 $\equiv$ parametric amplifier pump amplitude.}
\end{figure}


\begin{thebibliography}{99}

\bibitem{ben93} C.~H.~Bennett, G.~Brassard, 
C.~Crepeau, R.~Jozsa, A.~Peres 
and W.~K.~Wootters, Phys.Rev.Lett., {\bf70}, 1895 (1993).

\bibitem{bou97} D.~Boumeester {\it et al}, Nature (London) {\bf390}, 
575 (1997). 

\bibitem{bos98} D.~Boschi, S.~Branca, F.~De Martini, L.~Hardy and 
S.~Popescu, Phys.Rev.Lett., {\bf80}, 1121 (1998).

\bibitem{vai94} L.~Vaidman, Phys.Rev.A. {\bf 49}, 1473 (1994).

\bibitem{bra981} S.~L.~Braunstein and H.~J.~Kimble, Phys.Rev.Lett., {\bf80}, 
869 (1998).

\bibitem{ral981} T.~C.~Ralph and P.~K.~Lam, to appear in Phys.Rev.Lett., 
(1998).

\bibitem{fur98} A.~Furusawa, J.~L.~Sorensen, S.~L.~Braunstein, C.~A.~Fuchs, 
H.~J.~Kimble and E.~S.~Polzik, Science, {\bf 282}, 706 (1998).

\bibitem{bras98} G.~Brassard, S.~L.~Braunstein and R.~Cleve, Physica 
D {\bf120}, 43 (1998).

\bibitem{ben95} C.~H.~Bennett, Phys.Today {\bf 48}, 24 October (1995).

\bibitem{bra982} S.~L.~Braunstein, Nature {\bf 394}, 47 (1998).

\bibitem{pol98} R.~E.~S.~Polkinghorne and T.~C.~Ralph, submitted to 
Phys.Rev.Lett. (1998).

\bibitem{cav81} C.~M.~Caves, Phys.Rev.D. {\bf 23}, 1693 (1981).

\bibitem{ein35} A.~Einstein, B.~Podolsky and N.~Rosen, Phys.Rev., 
{\bf47}, 777 (1935).

\bibitem{ou92} Z.~Y.~Ou, S.~F.~Pereira, H.~J.~Kimble, and K.~C.~Peng, 
Phys.Rev.Lett. {\bf 68}, 3663 (1992).

\bibitem{yam86} Y.~Yamamoto, S.~Machida and O.~Nilsson, Phys.Rev.A. {\bf 
34}, 4025 (1986).

\bibitem{tsu98} M.~Tsurekane, S.~Kimura, M.~Kimura, N.~Taguchi and 
H.~Inaba, Appl.~Phys.~Lett {\bf 35}, 3586 (1985).

\bibitem{bra96} S.~L.~Braunstein, Phys.Rev.A. {\bf 72}, 3414 (1996).

\end{thebibliography}
\end{document}